\documentclass[aps,pra,twocolumn,superscriptaddress,showpacs]{revtex4-2}
\usepackage{amsmath,amssymb,wasysym,graphicx}
\usepackage[utf8]{inputenc}
\usepackage[T1]{fontenc}
\usepackage{xcolor}

\IfFileExists{newtxtext.sty}
{\usepackage{newtxtext,newtxmath}}
{\IfFileExists{stix.sty}
	{\usepackage{stix}}
	{\IfFileExists{mathptmx.sty}
		{\usepackage{mathptmx}}{} } }

\usepackage{textcomp}

\usepackage{bm}

\usepackage{ragged2e}
\usepackage{blindtext}

\IfFileExists{siunitx.sty}{\usepackage{booktabs,siunitx}}{}

\usepackage{color}
\definecolor{LinkColor}{rgb}{0.256,0.439,0.588}
\usepackage{hyperref}
\hypersetup{
	pdfauthor={good guys},
	pdftitle={good title},
	colorlinks=true,
	citecolor=LinkColor,
	linkcolor=LinkColor,
	urlcolor=LinkColor
}

\newcommand{\ket}[1]{\left|#1\right>}

\usepackage{latexsym}
\usepackage{amssymb}
\usepackage{amsmath}
\usepackage{amsfonts}
\usepackage{wasysym}
\usepackage[vcentermath]{youngtab}
\usepackage{upgreek}
\usepackage{bm,dsfont}
\usepackage{multirow}
\usepackage{enumitem}
\usepackage{makecell}
\usepackage{framed}
\usepackage{color}
\usepackage{hyperref}
\usepackage{tikz}
\hypersetup{
colorlinks = true,
linkcolor = [rgb]{0.70,0.13,0.13},
citecolor = [rgb]{0.13,0.55,0.13},
urlcolor = [rgb]{0.25, 0.41, 0.88}}

\def\be{\begin{equation}}
\def\ee{\end{equation}}
\newcommand{\beqn}{\begin{eqnarray}}
\newcommand{\eeqn}{\end{eqnarray}}

\begin{document}

\title{Topological edge states at Floquet quantum criticality}

\author{Longwen Zhou}
\email{zhoulw13@u.nus.edu}
\affiliation{College of Physics and Optoelectronic Engineering, Ocean University of China, Qingdao 266100, China}
\affiliation{Institute of Theoretical Physics, Chinese Academy of Sciences, Beijing 100190, China}

\author{Jiangbin Gong}
\email{phygj@nus.edu.sg}
\affiliation{Department of Physics, National University of Singapore, Singapore 117551, Singapore}
\affiliation{Centre for Quantum Technologies, National University of Singapore, Singapore 117543, Singapore}

\author{Xue-Jia Yu}
\email{xuejiayu@fzu.edu.cn}
\affiliation{Department of Physics, Fuzhou University, Fuzhou 350116, Fujian, China}
\affiliation{Fujian Key Laboratory of Quantum Information and Quantum Optics,
College of Physics and Information Engineering,
Fuzhou University, Fuzhou, Fujian 350108, China}

\begin{abstract}
Topologically protected edge states exactly at topological phase boundaries challenge the conventional belief that topological states must be associated with a bulk energy gap.  Because periodically driven (Floquet) systems host unusually intricate topological phase boundaries,  topological edge states can be prolific at such Floquet quantum criticality. Working on a class of chiral-symmetric, Floquet-driven Majorana fermion chains, we analytically and computationally show that the precise boundaries between different Floquet topological gapped phases can accommodate topological edge modes, including the so-called Majorana $\pi$ modes.    We also identify a general bulk-edge correspondence formula to predict and understand the emergence of topological edge modes at Floquet quantum criticality. Of direct interest to quantum simulation experiments, our results break new grounds for studies of nonequilibrium topological phases of matter undergoing topological phase transitions.


\end{abstract}

\maketitle

Topological phases form a cornerstone of modern condensed matter physics, extending beyond the Landau-Ginzburg-Wilson paradigm of symmetry-breaking order. A notable example of topological states is symmetry-protected topological (SPT) phases~\cite{hasan2010rmp,qi2011rmp,wen2019choreographed}, typically associated with a bulk energy gap.  It is widely believed that topological properties of SPT phases are destroyed when the bulk gap closes. However, recent progresses have shown that many key features of topological physics, such as topological edge modes, can surprisingly coexist with bulk critical fluctuations. This recognition has led to the definition of nontrivial topology at quantum critical points (QCPs), now known as topologically nontrivial QCPs or gapless SPT phases~\cite{fidkowski2011prb,kestner2011prb,keselman2015prb,ruhman2017prb,scaffidi2017prx,MAC3,parker2018prb,MAC4,verresen2021prx,thorngren2021prb,DuquePRB2021,friedman2022prb,yu2022prl,wen2023prb,yu2024prl,li2024sci_post,su2024prb,zhang2024pra,zhong2024pra}. The discovery of topologically protected edge states with gap closing suggests that topology still plays a crucial role in classifying phase transitions and in further detecting phase transitions along topological phase boundaries.  This opens a new avenue in studies of quantum phase transitions.

Nonequilibrium systems, particularly relevant to quantum simulation experiments~\cite{blatt2012quantum,gross2017quantum,browaeys2020many,kjaergaard2020superconducting} exhibit a number of intriguing topological phenomena absent in equilibrium counterparts.  As a prime example of nonequilibrium setting, periodically driven (Floquet) systems have realized a variety of exotic states of matter~\cite{oka2019floquet,else2020discrete,RODRIGUEZVEGA2021168434}.  Floquet systems hence offer a new stage in exploring novel physics including Floquet topological phases~\cite{GongPRA2008,OkaPRB2009,JiangPRL2011,DerekPRL2012,RudnerPRX2013,STF,TorresPRL2014,TianPRL2014,else2016prb,TenfoldF,lee2018prl,ZhouPRA2018,rodriguez-vega2018prl,ZhouPRB2019,WimbergerPRA2022,Zhao2022PRB,KQWZExp,ESEEFTP1}, Floquet quantum criticality and conformal field theory~\cite{khemani2016prl,berdanier2017prl,berdanier2018floquet,yates2018prl,han2020prb,wen2021prr,fan2021sci_post,Qin2022prb}, Floquet many-body localization~\cite{ponte2015prl,zhang2016prb,decker2020prl} and discrete time crystals~\cite{else2016prl} etc. Of particular importance is the following question. Can topologically protected edge states, which might not have static analogs in the first place, still coexist with Floquet gap closing (hence Floquet quantum criticality)?

To answer the question above, in this work we analytically solve the existence of topological edge modes, away or precisely on topological phase boundaries, for a broad class of free Majorana fermion chains~\cite{KC} under periodic driving.  Intriguingly, we discover that the topological phase diagrams of this class of models share a common structure.  This allows us to find systems with arbitrarily many Majorana edge modes appearing not only in gapped phases but also persisting along topological phase boundaries (i.e., coexisting with a gapless Floquet bulk spectrum). The topological phase boundaries not only host the familiar Majorana zero modes but also feature a new type of Majorana edge modes that are {\it absent in static situations}.  With a set of generalized winding numbers, we are able to identify a simple rule reflecting the bulk-edge correspondence at Floquet quantum criticality. In addition to providing several examples of lattice simulations to verify our analytical predictions, we have also computationally extended our general findings to nonequilibrium systems of $(2+1)$-dimensions.

\begin{figure}[tb]
    \includegraphics[width=\columnwidth]{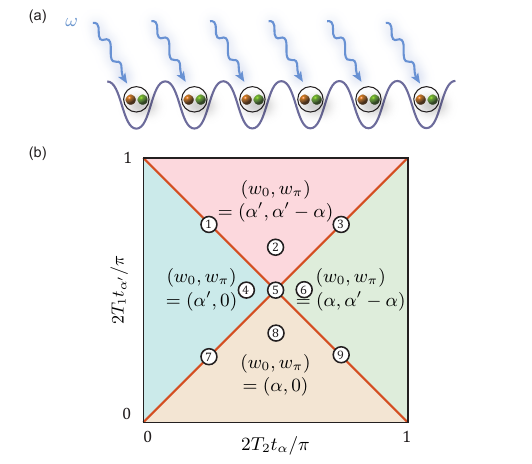}
    \caption{(a) A schematic plot of the Floquet-driven Majorana $\alpha$-chain. The blue wavy lines represent periodic drivings with frequency $\omega$. (b) The universal phase diagram of Eq.~(\ref{E1}), showing four gapped Floquet phases with each characterized by a pair of topological winding numbers $(w_{0}, w_{\pi}) \in \mathbb{Z} \times \mathbb{Z}$. Transitions between these gapped phases correspond to either topologically trivial or nontrivial Floquet quantum critical lines, as described in the main text. The numbered circles indicate representative locations within each phase and at phase transitions, corresponding to the quasienergy level diagram in Fig.~\ref{fig:ill}}
    \label{fig:phasediagram}
\end{figure}

\section*{Results}
\subsection*{Topological phase diagram of a class of Floquet Majorana chains}
Floquet topological phase boundaries can be extremely rich and this fact provides a promising opportunity to explore the existence of topological edge modes exactly at the topological phase boundaries. To that end, we investigate a class of Floquet-driven Majorana $\alpha$-chain (FMAC) at half-filling, described by the Hamiltonian
\begin{equation}
\label{E1}
    H(t)=\begin{cases}
H_{\alpha'}=-i\sum_{n}t_{\alpha'}\widetilde{\gamma}_{n}\gamma_{n+\alpha'}, & t\in[0,T_{1}),\\
H_{\alpha}=-i\sum_{n}t_{\alpha}\widetilde{\gamma}_{n}\gamma_{n+\alpha}, & t\in[T_{1},T_{1}+T_{2}),
\end{cases}
\end{equation}
where $\gamma_{n}$ and $\widetilde{\gamma}_{n}$ denote Majorana fermions at odd and even sublattice sites [marked as the orange and green balls in Fig.~\ref{fig:phasediagram}(a)]. The hopping strength $t_{\alpha^{\prime}(\alpha)}$ is nonzero for only a finite value of $\alpha^{\prime}(\alpha)$. For simplicity, we set the time unit as $T=T_1+T_2=1$ and assume $\alpha^{\prime} > \alpha$ in the following discussion (similar results hold for $\alpha^{\prime} < \alpha $). The Hamiltonian $H(t) = H(t+T)$ is time-periodic and the eigenstates of its propagator $F=\hat{\mathsf{T}}e^{-i\int_0^T H(t)dt}$ are Floquet states $\ket{\psi(t)} = e^{-i E t} \ket{\Phi(t)}$, where $\hat{\mathsf{T}}$ is the time ordering operator, $\ket{\Phi(t+T)} = \ket{\Phi(t)}$, in analogy with Bloch's theorem for systems with discrete space-translational symmetry. Consequently, the energy is replaced by quasienergy $E$, which is defined modulus $2\pi$ and only conserved stroboscopically. Additionally, $F$ exhibits the chiral symmetry, enabling the definition of winding numbers to characterize its nontrivial topology [for details, see Secs.~I--III of the Supplemental Materials (SM)].

As one static case for comparison, the model $H_{\rm s}=H_\alpha+H_{\alpha'}$ could undergo a transition between topologically distinct gapped phases, characterized by a stack of Majorana conformal field theories (CFTs) with central charge $c=(\alpha^{\prime}-\alpha)/2$~\cite{MAC3}. These phases and transitions are topologically nontrivial when $\alpha>0$, as both feature exponentially localized Majorana zero modes. The exact number of these modes can be analytically determined by a single topological winding number that depends on $\alpha$ and $\alpha'$ \cite{MAC3}. However, back to the current nonequilibrium system periodically quenched between  $H_\alpha$ and $H_{\alpha'}$,  the corresponding topological phase diagram is obtained and shown in Fig.~\ref{fig:phasediagram}(b). As seen from Fig.~\ref{fig:phasediagram}(b), there are  now four gapped Floquet phases, leading to multiple topological phase boundaries.  Each gapped Floquet phase is characterized by a pair of topological winding numbers ($w_0, w_{\pi}$). The first one, $w_0$, counts the number of Majorana edge modes at quasienergy $0$, which are already present in the static system $H_{\rm s}$ mentioned above. In contrast, $w_{\pi}$ counts the second type of Majorana edge mode at quasienergy $\pi$ (a feature unique to Floquet settings),  which arises from chiral plus discrete-time translational symmetries. The boundaries between different Floquet phases can be determined by the condition $2T_{1}t_{\alpha'}\pm2T_{2}t_{\alpha}=\nu\pi,\nu\in\mathbb{Z}$ (for details, see Sec.~II of the SM). Such phase boundaries are a result of Floquet gap-closing at either $0$ or $\pi$ quasienergy. Accordingly, the central charge incorporating both gap-closing scenarios is found to be $c=(\alpha^{\prime}-\alpha)/2$ on any normal phase boundary and $c=\alpha^{\prime}-\alpha$ for the multicritical point as the intersection of two topological phase boundaries  (see Secs.~II and VIII of the SM for details).

\subsection*{Topological edge states precisely on topological phase boundaries}
Our nonequilibrium model allows us to solve exactly the Majorana edge modes associated with different topological phases with a Floquet quasienergy gap, but also offers opportunities to inspect the fate of such topological modes when the Floquet quasienergy gap closes at phase boundaries. Indeed, it is possible to analytically show that for any $\alpha>0$ and $\alpha'\neq\alpha$, topologically protected edge modes do sustain the closure of a Floquet quasienergy gap. Notably, the number and precise expressions of these Majorana edge modes on Floquet phase boundaries can be found by first extending and then inspecting the same expressions (e.g., checking localization and normalization features) for topological edge modes with a nonzero quasienergy gap to cases with vanishing quasienergy gap. As an explicit example, a half-infinite chain with $(\alpha,\alpha')=(1,2)$ (see Sec.~V of the SM) is found to yield the following exact wavefunctions for two Majorana zero modes 
\begin{alignat}{1}
\gamma_{\rm L}^{(0,1)}&=  \gamma_{1}\nonumber \\
\gamma_{\rm L}^{(0,2)}&=  \sum_{n=1}^{\infty}\left[-\frac{\tan(T_{2}t_{1})}{\tan(T_{1}t_{2})}\right]^{n-1}\left[\sin(T_{2}t_{1})\widetilde{\gamma}_{n}-\cos(T_{2}t_{1})\gamma_{n+1}\right],
\end{alignat}
and the following wavefunction for one Majorana $\pi$ mode
\begin{alignat}{1}
\gamma_{\rm L}^{(\pi)}= & \sum_{n=1}^{\infty}\left[\frac{1}{\tan(T_{2}t_{1})\tan(T_{1}t_{2})}\right]^{n-1}\nonumber \\
\times & \left[\cos(T_{2}t_{1})\widetilde{\gamma}_{n}+\sin(T_{2}t_{1})\gamma_{n+1}\right].
\end{alignat}
Here L indicates being at the left edge of the half-infinite chain.  With these explicit expressions, one can easily verify their existence at Floquet quantum criticality.

Two general observations can be made first.  Firstly, for $\alpha=0$ and $\alpha'>0$, the static Majorana chain $H_{\rm s}$ does not host topological edge modes when tuning the system to its topological phase boundaries. However, periodic driving can still induce new topological phase boundaries that will host topological edge modes. Secondly, for $\alpha^{\prime} > \alpha > 0$, although the topological phase boundaries of the periodically quenched quantum chain may host topological edge states, the nonequilibrium setting induces additional topologically nontrivial quantum phase boundaries. Importantly, these newly formed topological phase boundaries not only host the Majorana zero modes but also accommodate Majorana $\pi$ modes, with the latter being unique to Floquet settings.

\begin{figure}[tb]
    \includegraphics[width=\columnwidth]{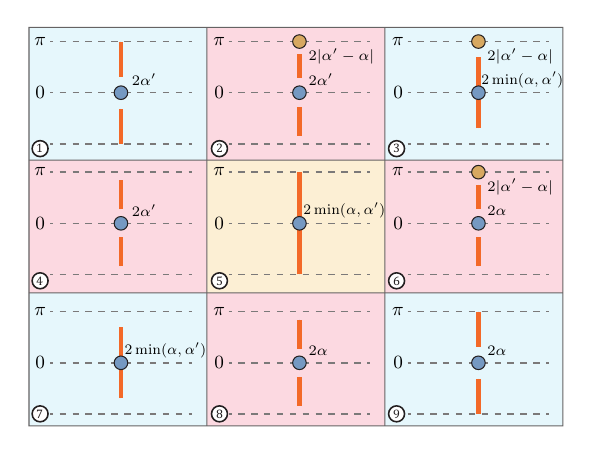}
    \caption{Schematic quasienergy level diagram of each phase and phase transition. The deep blue and yellow dots represent Majorana edge modes at quasienergies 0 and $\pi$. The number next to each dot indicate the exact number of Majorana edge modes, which can be determined analytically (see Secs.~III and IV of the SM for details). The numbered circles correspond to the representative points in Fig.~\ref{fig:phasediagram}}
    \label{fig:ill}
\end{figure}

To help to digest our key results qualitatively, we present a schematic quasienergy-level diagram in Fig.~\ref{fig:ill}. In each phase (marked by \textcircled{2},\textcircled{4},\textcircled{6},\textcircled{8}), two bulk gaps exist at quasienergies $0$ and $\pi$, allowing for hosting Majorana zero (deep blue dots) or $\pi$ (deep yellow dots) edge modes under open boundary conditions, whose exact numbers depend on $\alpha$ and $\alpha'$. Along the topological phase boundaries (marked by \textcircled{1},\textcircled{3},\textcircled{7},\textcircled{9}), the two-gap structure afforded by our nonequilibrium system results in the bulk gap closing at one quasienergy ($0$ or $\pi$), whereas the Majorana edge modes remain protected by the other gap at a different quasienergy (for examples of Majorana edge modes coexisting with a fully gapless bulk, see Sec.~VI of the SM). This mechanism explains why periodic driving can lead to richer topologically nontrivial phase boundaries. At the multicritical point (marked by \textcircled{5}), both bulk quasienergy gaps vanish, yielding Majorana zero modes localized at the boundary for any $\alpha>0$.

Based on our general remarks and a qualitative summary above, we are now ready to elaborate on specific cases below. 

\subsection*{Majorana $\pi$ edge modes at Floquet quantum criticality: $(\alpha,\alpha')=(0,1)$}
Though our nonequilibrium models can be solved analytically, it is helpful to computationally verify the analytic predictions using
specific examples. As the first example, we consider the model with $(\alpha,\alpha')=(0,1)$, which can be mapped to the well-known transverse field Ising model via the Jordan-Wigner transformation. In the static case,  $H_{\rm s}$ exhibits magnetic long-range order and trivial paramagnetic (PM) phases, corresponding to topological and trivial phases in the fermionic language. The topological phase transition point at $t_0=t_1$ is topologically trivial, hence yielding no topologically protected zero modes at this phase boundary.

With periodic driving, the transverse field Ising model remains solvable in the Majorana representation (see Sec.~II of the SM for details). Without loss of generality, we set $T_1=T_2=T/2$ in the following calculations. The resulting phase diagram includes two familiar phases: the PM and spin glass (SG) phases that inherit from the static counterpart $H_{\rm s}$, {\it plus}  two new phases: the $\pi$-SG and 0$\pi$ PM (a Floquet SPT) phases \cite{khemani2016prl}, unique to nonequilibrium systems. We note a subtle difference in the phase transitions between the PM phase and the two SG phases, due to gap-closing at quasienergy $0$ or $\pi$ in the fermionic language. The boundaries dividing the Floquet SPT phase and the two SG phases (boundaries labeled by \textcircled{7} and \textcircled{9} in Fig.~\ref{fig:phasediagram}) are topologically nontrivial, representing a special type of Floquet gapless SPT ``phase''. Furthermore, there are two distinct types of topologically nontrivial Floquet phase boundaries: one hosts Majorana zero modes (boundary between the SG phase and the $0\pi$ PM phase, labeled by \textcircled{1} in Fig.~\ref{fig:phasediagram}) and shares the same topology as in the static counterpart, whereas the other features Majorana edge modes at quasienergy $\pi$ (boundary between $\pi$-SG phase and  the $0\pi$ PM phase, labeled by \textcircled{3} in Fig.~\ref{fig:phasediagram}). Such Majorana $\pi$ edge modes albeit Floquet gap closure are protected by an emergent symmetry inherited from discrete-time translation symmetry and have no analog in equilibrium systems.  Evidently, even when starting from a topologically trivial static Hamiltonian, periodic driving can induce topologically nontrivial phase boundaries hosting Majorana $\pi$ edge modes. It is important to emphasize that a Floquet topological semimetal~\cite{bomantara2016pre,hubener2017creating,Li2018prl} can be considered a gapless SPT phase in a broader sense. The key distinction between gapless SPT phases and our findings here lies in their topological properties: the former relies on space-translational symmetry, making it susceptible to destabilization by disorder. In contrast, the topological edge modes at Floquet quantum criticality remains robust even in the presence of symmetry-preserving disorder.

\subsection*{Topological phase transitions at Floquet quantum criticality: $(\alpha,\alpha')=(1,2)$}
\begin{figure}[tb]
    \includegraphics[width=\columnwidth]{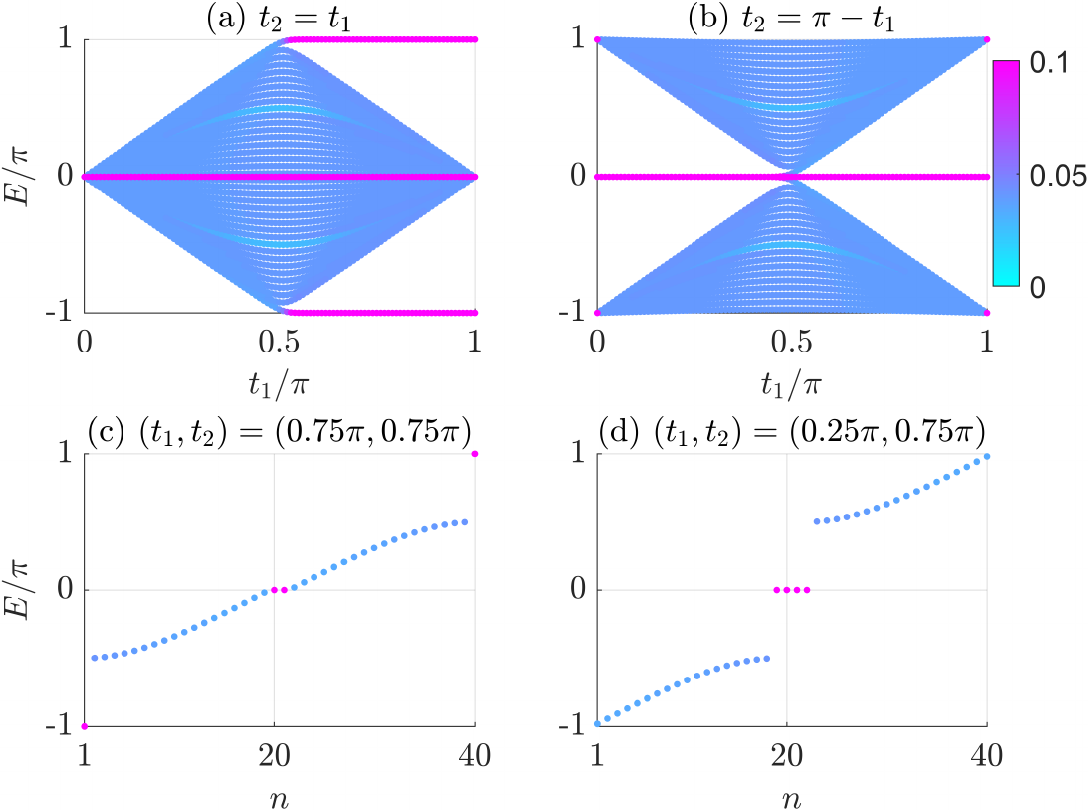}
    \caption{Floquet quasienergy spectra under the OBC along quantum critical lines $t_1 = t_2$ (a) and $t_1 = \pi - t_2$ (b). (c) At $(t_1,t_2) = (0.75\pi,0.75\pi)$, the bulk spectrum is gapless at $E=0$ and gapped at $E = \pm \pi$. There are two degenerate Majorana edge modes at $E = 0$ and another two at $E=\pi$. (d) At $(t_1,t_2)=(0.25\pi,0.75\pi)$, the bulk spectrum is gapless at $E = \pi$ and gapped at $E = 0$, featuring four degenerate Majorana zero modes. All panels are computed with $20$ unit cells, i.e., $40$ sites in the Majorana basis. $E$ represents the quasienergy. $n$ is the state index. The inverse participation ratios of different Floquet states are represented by the color bar.}
    \label{fig:1d}
\end{figure}

As a minimal example to demonstrate intriguing and rich topological implications for phase boundaries, we return to the case of $(\alpha,\alpha')=(1,2)$ mentioned above.   The static counterpart hosts exponentially localized Majorana zero modes  at its gap-closing boundary, a fact related to symmetry-enriched conformal field theory explored in the cluster Ising model through duality mapping~\cite{verresen2021prx,yu2022prl}. Upon periodic driving that yields four gapped Floquet topological phases, there are now four topological phase boundaries that continue to host Majorana zero modes. Along the phase boundary $t_2 = t_1$ [Fig.~\ref{fig:1d}(a)], the quasienergy spectrum hosts Majorana modes that are either only pinned at zero quasienergy ($t_1\leq 0.5\pi$) or at both the zero and $\pi$ quasienergies ($t_1 > 0.5 \pi$). These specific results are particularly interesting because we are observing topological phase transitions along the topological phase boundaries!  
In particular, the coexistence of Majorana zero and $\pi$ modes [see Fig.~\ref{fig:1d}(c)] at Floquet quantum criticality is a topological phenomenon unique to nonequilibrium systems.  This indicates that even double-period (``$2T$'') time-crystal like oscillations arising from a coherent superposition of Majorana $0$ and $\pi$ modes \cite{GongPRL18} can sustain Floquet quasienergy gap closing.   
Furthermore, as shown in Fig.~\ref{fig:1d}(b), the two phase boundaries along $t_2 + t_1 = \pi$ before or after passing the multicritical point $(t_1=t_2=0.5\pi)$ both host Majorana zero modes only. The difference between these two phase boundaries is that the former hosts two pairs of Majorana modes (as indicated in Fig.~\ref{fig:1d}(d) and Sec.~V of the SM), whereas the latter hosts only one pair.   This is again an intriguing example of topological phase transitions at Floquet quantum criticality. Loosely speaking, we have thus uncovered a new mechanism for \textit{phase transitions of (or ``higher-order'') topological phase transitions}.

\subsection*{Bulk-edge correspondence for topological edge states at Floquet quantum criticality}
Topological edge states at Floquet phase boundaries can be protected by the emergent symmetry inherited from discrete-time translation symmetry, analogous to symmetry-enriched quantum criticality in equilibrium systems~\cite{verresen2021prx}.  The underlying nontrivial topology at Floquet quantum criticality can be further investigated by the entanglement spectrum under periodic boundary condition~\cite{ESEEFTP1}, which serves as a bulk diagnostic that is insensitive to boundary conditions, as illustrated in Sec.~VIII of the SM. 

Based on extensive computational and analytical treatments of our nonequilibrium model, we are able to identify a general bulk-edge correspondence rule to account for the existence and the precise number of topological edge modes at Floquet quantum criticality  for any $\alpha^{\prime} \neq \alpha$ (detailed in Secs.~III and IV of the SM), namely, 
\begin{equation}
\label{E2}
(N_0,N_{\pi}) = 2(|\omega_0|,|\omega_{\pi}|).
\end{equation}
This general correspondence indicates that at the Floquet phase boundaries, the numbers of Majorana zero and $\pi$ edge modes, $(N_{0}, N_{\pi})$, are entirely determined by the bulk generalized winding numbers $(\omega_0,\omega_{\pi})$. Because the winding numbers $\omega_0$ and $\omega_{\pi}$ can be arbitrarily large by setting different values of $\alpha$ and $\alpha'$,  it is clear that  topological properties of nonequilibrium systems at gap-closing boundaries are much richer than their static counterparts. 
In particular, for systems with larger values of $\alpha$ and $\alpha'$, more intriguing examples of topological phase transitions precisely at Floquet gap-closing phase boundaries can be identified (see Sec.~VI of the SM for details).

\begin{figure}[tb]
    \includegraphics[width=\columnwidth]{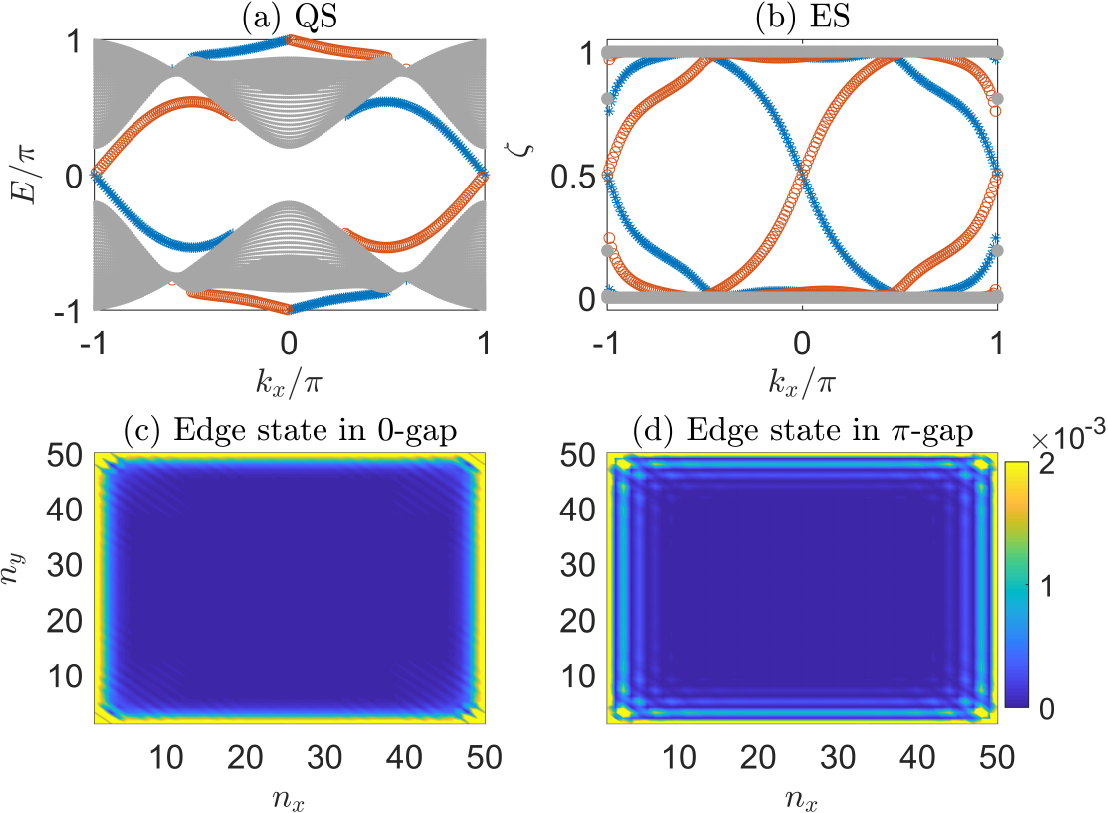}
    \caption{An illustration of topologically nontrivial Floquet quantum criticality in 2D with the kicked QWZ model. (a) Floquet quasienergy spectrum (QS) $E(k_x)$ under the PBC (OBC) along $x$ ($y$) direction. Gray dots, blue stars and red circles denote bulk states, left-localized edge states and right-localized edge states. The bulk Floquet bands touch at $E=\pi$. (b) Entanglement spectrum (ES) $\zeta(k_x)$ under the PBC, with entanglement cuts taken along $y$. Gray dots, blue stars and red circles denote extended, left-localized and right-localized eigenmodes of the Floquet entanglement Hamiltonian. (c) Probability distribution of a typical chiral edge state crossing the $E=0$ quasienergy gap. (d) Probability distribution of a chiral edge state at $E=\pi$. (c) and (d) share the same color bar.}
    \label{fig:2d}
\end{figure}

\subsection*{Topological edge modes at Floquet quantum criticality in (2+1) dimensions}
We investigate topological edge modes at Floquet quantum criticality in $(2+1)$ dimensions, extending beyond the general framework of the one-dimensional chiral symmetric model studied above. As a concrete example, we consider the two-dimensional Qi-Wu-Zhang (QWZ) model with a periodically $\delta$-kicked mass term~\cite{QWZ,KQWZ1} (details can be found in Sec.~VII of the SM), termed the kicked QWZ (KQWZ) model.  The Floquet operator is given by $U(k_{x},k_{y})=\text{exp}(-iV\sigma_{z})\text{exp}(-iJ[\sin k_{x}\sigma_{x}+\sin k_{y }\sigma_{y}+(\cos k_{x}+\cos k_{y})\sigma_{z}])$ where $V$ and $J$ are dimensionless parameters. In the absence of Floquet driving, the topological phase boundary of the QWZ model hosts no topological edge modes. However, with periodic driving, the quasienergy spectrum can host two pairs of chiral edge states crossing $0$ and $\pi$ quasienergies, represented by the red circles and blue stars in Fig.~\ref{fig:2d}(a).  Notably, the number of critical edge states observed at the gap-closing boundary can also be inferred from the bulk entanglement spectrum, as depicted in Fig.~\ref{fig:2d}(b). Furthermore, these chiral edge states sustaining Floquet gap-closing can be directly traced in real space, with the wave function probability distributions shown in Figs.~\ref{fig:2d}(c) and \ref{fig:2d}(d) for quasienergies $0$ and $\pi$, respectively. These results provide compelling evidence that nonequilibrium setting can host a variety of topological edge modes at topological phase boundaries in systems of higher dimensions. 

\begin{figure}
\begin{centering}
\includegraphics[scale=0.48]{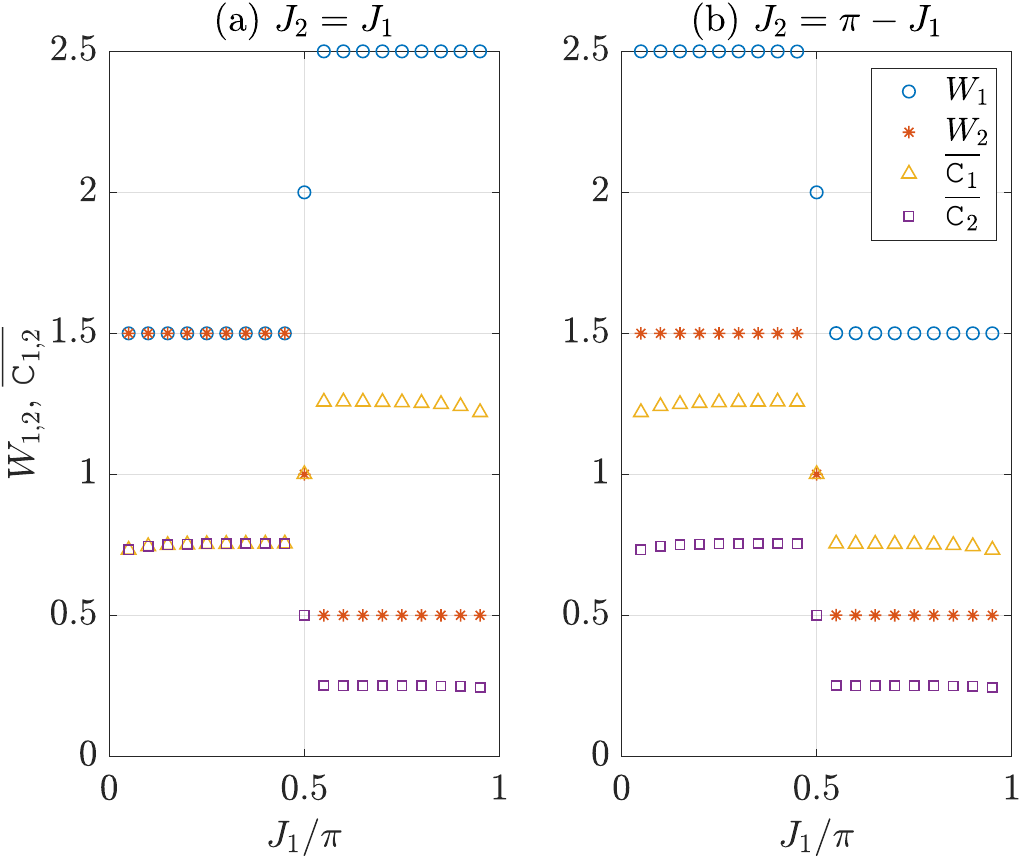}
\par\end{centering}
\caption{The DWNs $(W_{1},W_{2})$ and MCDs $(\overline{\mathsf{C}_{1}},\overline{\mathsf{C}_{2}})$
of FMAC with $(\alpha,\alpha')=(1,2)$ in two symmetric time frames
and along the critical lines (a) $J_{1}=J_{2}$ and (b) $J_{1}=\pi-J_{2}$.
Simplified notations $J_{1}\equiv2T_{2}t_{1}$ and $J_{2}\equiv2T_{1}t_{2}$
are used. All the data are obtained after averaging over $M=40$ driving
periods. In the long-time limit $M\rightarrow\infty$, we theoretically
expect $\overline{\mathsf{C}_{1}}+\overline{\mathsf{C}_{2}}=\frac{W_{1}+W_{2}}{2}=w_{0}$
and $\overline{\mathsf{C}_{1}}-\overline{\mathsf{C}_{2}}=\frac{W_{1}-W_{2}}{2}=w_{\pi}$,
which are consistent with the reported results of numerical simulations
up to finite-time errors. \label{fig:DWNMCD}}
\end{figure}

\begin{figure}
\begin{centering}
\includegraphics[scale=0.46]{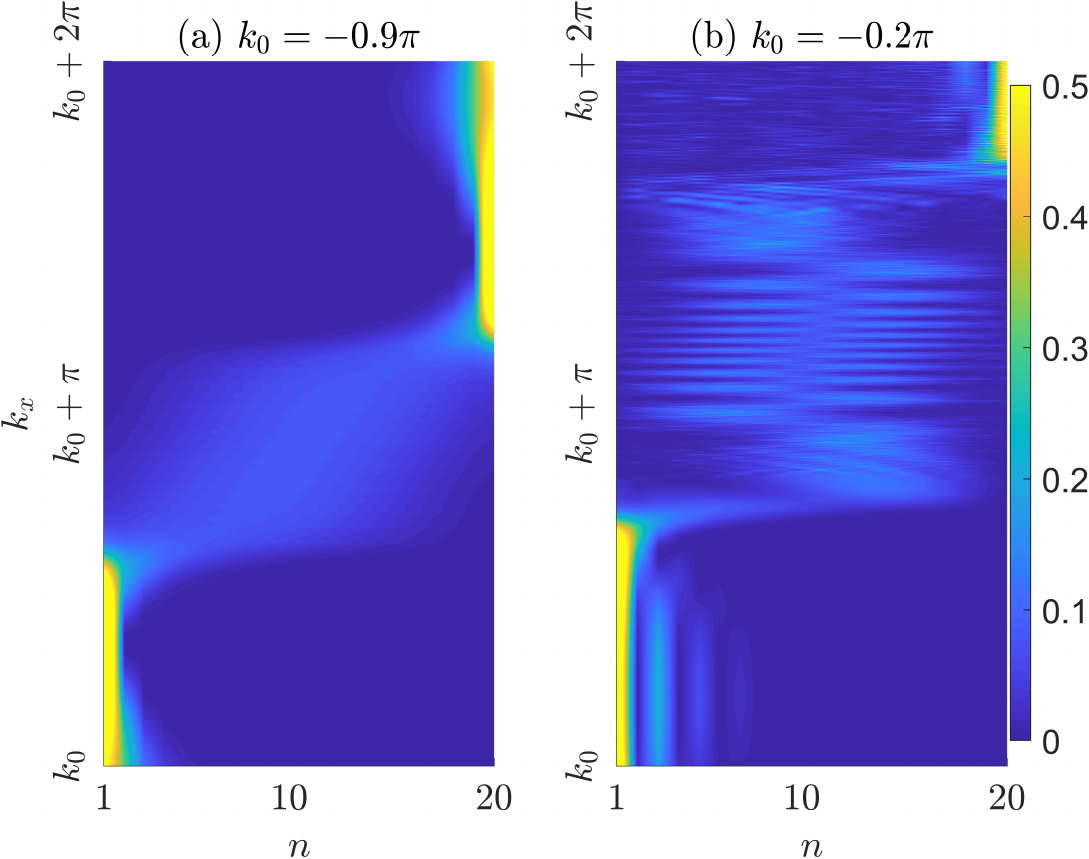}
\par\end{centering}
\caption{Adiabatic pumping of edge states in the KQWZ model. The PBC (OBC)
is taken along $x$ ($y$) direction. $n$ is the unit-cell index
along $y$. The color bar shows the probability distribution of the
pumped state at different $k_{x}$ in the lattice. Other system parameters
are $(J,V)=(0.6\pi,0.2\pi)$. Initial states are chosen as the left
edge states with $k_{x}=k_{0}=-0.9\pi$ and $-0.2\pi$ in (a) and
(b), respectively. Their quasienergies are around the gaps at $E=0$
and $E=\pm\pi$ in Fig.~10(c) of the SM. \label{fig:KQWZESP}}
\end{figure}

\subsection*{Possible observations of Floquet topology at quantum criticality}
We now consider possible experimental detection of topological states at Floquet gap-closing points using cold-atom-based quantum wires. In this context,
Floquet Majorana edge modes may be detected by spatially resolved
rf spectroscopy~\cite{JiangPRL2011}. The bulk topological invariants
may be extracted from the dynamic winding numbers
(DWN) of time-averaged spin textures (TAST) or the mean chiral displacements (MCD) \cite{LiuSciBul2018,LeePRR2020,CardanoNC2017}.


In Fig.~\ref{fig:DWNMCD}, we illustrate the usefulness
of DWN and MCD in detecting the topological invariants of FMAC with
the representative case $(\alpha,\alpha')=(1,2)$ (see Methods for details). We observe that
after a finite number of evolution periods, these quantities could
offer clear dynamical signatures for Floquet topological criticality
and quantum transitions between different Floquet gapless topological
phases. Therefore, for an arbitrary FMAC with $\alpha\neq\alpha'$,
we may detect its topological properties by measuring the DWN and
MCD in experiments.

For the 2D KQWZ model, we may perform edge state pumping \cite{KrausPRL2012}
to recognize the presence of chiral edge states at its Floquet quantum
criticality. The idea is to consider the system under the PBC along
one dimension and OBC along the other, and treating the quasimomentum
along the PBC-direction as an adiabatic parameter. For example, we
may take the PBC (OBC) along $x$ ($y$) direction for the KQWZ. In
this case, the Floquet operator is given by the $F(k_{x})$ in Eq.~(183) of the SM, and we could set the quasimomentum $k_{x}$ as the
adiabatic parameter. 

To provide further evidence, we simulate the edge state pumping computationally
using the KQWZ model, with results reported in Fig.~\ref{fig:KQWZESP} (see Methods for details).
A pumping cycle is divided into $M=4000$ steps, with $\delta k=2\pi/M$
so as to ensure the adiabaticity. We observe that in both cases, the
left edge states are pumped to the right edges after an adiabatic
cycle, even though the system parameters are set at the critical points.
Therefore, we may utilize the edge state pumping to reveal the presence
of chiral edge bands at the topological phase boundaries of Floquet
Chern insulators.

\section*{Discussion}
To summarize, using a broad class of Floquet Majorana $\alpha$-chain models with chiral symmetry, we have shown, both analytically and computationally, the general existence of topological edge modes precisely at topological phase boundaries where at least one Floquet quasienergy gap closes.  The number of such exotic edge modes is determined by a general bulk-edge correspondence rule, hence firmly connected with the bulk properties.  Floquet topological phase boundaries can not only host the familiar topological Majorana zero modes but also  Majorana $\pi$ modes.  We have shown that for $(2+1)$D systems in symmetry class A, chiral edge states can also sustain quasienergy gap closing, with solid computational evidence of bulk-edge correspondence.  Extending our findings to other symmetry classes of Floquet free-fermion models, and more importantly, to Floquet topological systems with interactions, is a fascinating but more involved task. We have also briefly discussed a protocol for experimentally probing the predicted topological states at Floquet quantum criticality. Our study shall stimulate future studies of gapless bulk topology vs the physics at the edge of nonequilibrium systems, including ``higher-order'' topological phase transitions along the already very rich Floquet topological phase boundaries.

\section*{Methods}
\subsection*{DWN and MCD}
The DWN is suitable for detecting topological invariants of chiral
symmetric Floquet systems in momentum space~\cite{ChenPR2021}. For
our 1D system, we denote the eigenstates of its Floquet operator $F_{j}(k)$
in symmetric time frames $j=1,2$ by $|\psi_{j}^{\pm}(k)\rangle$,
whose quasienergies $\pm E(k)$ are given by Eq.~(38) of the SM.
An arbitrary initial state may then be expressed as $|\psi_{j}(k)\rangle=c_{j}^{+}|\psi_{j}^{+}(k)\rangle+c_{j}^{-}|\psi_{j}^{-}(k)\rangle$,
where $|c_{j}^{+}|^{2}+|c_{j}^{-}|^{2}=1$ due to the normalization
condition. After the evolution over $m$ driving periods by $F_{j}(k)$
in the time frame $j$, we arrive at the state
$|\psi_{j}^{m}(k)\rangle\equiv F_{j}^{m}(k)|\psi_{j}(k)\rangle=e^{-iE(k)m}c_{j}^{+}|\psi_{j}^{+}(k)\rangle+e^{iE(k)m}c_{j}^{-}|\psi_{j}^{-}(k)\rangle$.
For any Pauli spin $\sigma_{s}$, we obtain its average over the state
$|\psi_{j}^{m}(k)\rangle$ as
$\langle\sigma_{s}(k)\rangle_{j}^{m}=\langle\psi_{j}^{m}(k)|\sigma_{s}|\psi_{j}^{m}(k)\rangle$, where $s=x,y,z$.
In the long-time limit, the stroboscopic TAST of our FMAC is given
by the configuration of $[r_{y}^{j}(k),r_{z}^{j}(k)]$ in $k$-space,
where $r_{s}^{j}(k)\equiv\lim_{M\rightarrow\infty}\frac{1}{M}\sum_{m=1}^{M}\langle\sigma_{s}(k)\rangle_{j}^{m}$ ($s=y,z$).
At a given quasimomentum $k\in[-\pi,\pi)$, the dynamical winding
angle of stroboscopic TAST is defined as
$\theta_{j}(k)\equiv\arctan[r_{z}^{j}(k)/r_{y}^{j}(k)]$ ($j=1,2$).
It can be shown that the winding angle of stroboscopic TAST around
the first BZ $k\in[-\pi,\pi)$ generates the DWN $W_{j}\equiv\int_{-\pi}^{\pi}\frac{dk}{2\pi}\partial_{k}\theta_{j}(k)$ ($j=1,2$)
\cite{ZhouPRB2019}.
Moreover, it can be proved that \cite{ZhouPRB2019} so long as the
amplitudes of initial state $|\psi_{j}(k)\rangle$ satisfy $|c_{j}^{+}|\neq|c_{j}^{-}|$,
we would have $(W_{1},W_{2})=(w_{1},w_{2})$, where $w_{1,2}$ are
given by the Eq.~(45) of the SM. Therefore, by measuring the stroboscopic
TAST and extracting the DWNs $(W_{1},W_{2})$, one can detect the
bulk topological invariants $(w_{0},w_{\pi})$ of the FMAC defined in Eq.~(47) of the SM through the relations
\begin{equation}
(w_{0},w_{\pi})=\frac{1}{2}(W_{1}+W_{2},W_{1}-W_{2}).\label{eq:w0pW12}
\end{equation}

The MCD is suitable for detecting topological invariants of chiral
symmetric Floquet systems in real space \cite{WimbergerPRA2022}.
Let $F_{j}$ ($j=1,2$) be the Floquet operator of our FMAC in the
symmetric time frame $j$. Using the BdG basis of complex fermions
to express $F_{j}$ in real space, we find the MCD after $m$ evolution
periods as $\mathsf{C}_{j}(m)=\langle\psi_{0}|F_{j}^{-m}(\hat{n}\otimes{\cal S})F_{j}^{m}|\psi_{0}\rangle$.
Here, $\hat{n}$ is the position operator of unit cells and ${\cal S}$
is the chiral symmetry operator, which is equal to $\sigma_{x}$ in
the BdG basis. The initial state $|\psi_{0}\rangle$ at $t=0$ can
be chosen as one eigenstate of $\sigma_{x}$ in the ($n=0$)-position
sector. Under the PBC, it can be shown that $\mathsf{C}_{j}(m)=\frac{w_{j}}{2}-\int_{-\pi}^{\pi}\frac{dk}{2\pi}\frac{\cos[E(k)m]}{2}\partial_{k}\phi_{j}(k)$ \cite{ZhouPRA2018},
where $w_{j}$ is the winding number of $F_{j}(k)$, $E(k)$ is the
quasienergy of $F_{j}(k)$, and $\phi_{j}(k)$ is given by Eq.~(46) in the SM
for the FMAC. In terms of $\mathsf{C}_{j}(m)$, we can find the time-averaged
MCD over $m$ driving periods as
$\overline{\mathsf{C}_{j}(m)}\equiv\frac{1}{mT}\sum_{m'=1}^{m}\mathsf{C}_{j}(m)$.
It can be further proved that in the long-time limit $m\rightarrow\infty$,
we have $\overline{\mathsf{C}_{j}(m)}\rightarrow w_{j}/2$ \cite{ZhouPRA2018}.
Therefore, the topological invariants $(w_{0},w_{\pi})$ of FMAC can
be extracted dynamically from the time-averaged MCD through the following
relations
\begin{equation}
(w_{0},w_{\pi})=\lim_{m\rightarrow\infty}(\overline{\mathsf{C}_{1}(m)}+\overline{\mathsf{C}_{2}(m)},\overline{\mathsf{C}_{1}(m)}-\overline{\mathsf{C}_{2}(m)}).\label{eq:w0pC12}
\end{equation}
Similar to the DWN, these relations also hold for both the gapped
and critical phases of our FMAC.

\subsection*{Edge state pumping}
Initially, we can prepare the system at an edge state
$|\psi(k_{x}=k_{0})\rangle$, whose quasienergy could be around either
$E=0$ or $E=\pm\pi$. Next, we evolve the system over many driving
periods. The adiabatic parameter $k_{x}$ remains unchanged within
each driving period and increases by a small amount $\delta k$ at
the start of the next driving period~\cite{DerekPRL2012}. In such
a dynamical process, the initial state evolves according to
$|\psi(k_{0}+M\delta k)\rangle= F(k_{0}+M\delta k)F(k_{0}+(M-1)\delta k)\cdots F(k_{0}+2\delta k)F(k_{0}+\delta k)|\psi(k_{0})\rangle$
after a number of $M$ driving periods. When $M\delta k=2\pi$, the
adiabatic parameter has gone over a cycle, as $k_{x}$ is $2\pi$-periodic.
Stroboscopically, the evolution reaches the adiabatic limit when $\delta k=2\pi/M\rightarrow0$.
Our system could now undergo a cyclic adiabatic evolution with the
change of $k_{x}$. Importantly, if $F(k_{x})$ has a pair of chiral
edge bands and the initial state is prepared at one edge, it will
be pumped to the opposite edge during an adiabatic cycle $k_{x}=k_{0}\rightarrow k_{0}+2\pi$ \cite{KrausPRL2012}.
With this physics of edge-state pumping, we may then experimentally verify the existence of chiral edge
states in the KQWZ model with quasienergy gap closing. 

\textit{Acknowledgment}: We thank Xueda Wen, Chinghua Lee, Hongzheng Zhao, Chushun Tian, and Romain Vasseur for helpful discussions. 
L.~Zhou is supported by the National Natural Science Foundation of China (Grants No.~12275260, No.~12047503, and No.~11905211), the Fundamental Research Funds for the Central Universities (Grant No.~202364008), and the Young Talents Project of Ocean University of China.
X.-J.~Yu is supported by the National Natural Science Foundation of China (Grant No.~12405034). J.G. acknowledges support by the National
Research Foundation, Singapore and A*STAR under its CQT Bridging Grant.

\textit{Note added.} 
After our first draft was posted on arXiv, we noticed a related work arXiv:2411.02526, which appeared two weeks later and provided strong support for our conclusions.

\bibliography{main}

\end{document}